# RECOVERY OF 150-250 MeV/nuc COSMIC RAY HELIUM NUCLEI INTENSITIES BETWEEN 2004-2010 NEAR THE EARTH, AT VOYAGER 2 AND AT VOYAGER 1 IN THE HELIOSHEATH – A TWO ZONE HELIOSPHERE

**W.R Webber[1], F.B. McDonald[2], P.R. Higbie[3] and B. Heikkila[4]**

1. New Mexico State University, Department of Astronomy, Las Cruces, NM, 88003, USA
2. University of Maryland, Institute of Physical Science and Technology, College Park, MD, 20742, USA
3. New Mexico State University, Physics Department, Las Cruces, NM, 88003, USA
4. NASA/Goddard Space Flight Center, Greenbelt, MD 20771, USA

## Abstract

The recovery of cosmic ray He nuclei of energy ~150-250 MeV/nuc in solar cycle #23 from 2004 to 2010 has been followed at the Earth using IMP and ACE data and at V2 between 74-92 AU and also at V1 beyond the heliospheric termination shock (91-113 AU). The correlation coefficient between the intensities at the Earth and at V1 during this time period is remarkable (0.921), after allowing for a ~0.9 year delay due to the solar wind propagation time from the Earth to the outer heliosphere. The intensity measured at V1 is ~6 times that at the Earth in 2005 at the beginning of the recovery but at 2010.5 this difference is only a factor ~2.2 as a result of the fact that the relative intensity increase at the Earth is larger than that at V1. To describe these intensity changes and to predict the absolute intensities measured at all three locations we have used a simple spherically symmetric (no drift) two-zone heliospheric transport model with specific values for the diffusion coefficient in both the inner and outer zones. The diffusion coefficient in the outer zone, assumed to be the heliosheath from about 90 to 120 (130) AU, is determined to be ~5 times smaller than that in the inner zone out to 90 AU. This means the Heliosheath acts much like a diffusing barrier in this model. The absolute magnitude of the intensities and the intensity changes at V1 and the Earth are described to within a few percent by a diffusion coefficient that varies with time by a factor ~4 in the inner zone and only a factor of ~1.5 in the outer zone over the time period from 2004-2010. For V2 the observed intensities follow a curve that is as much as 25% higher than the calculated intensities at the V2 radius and at times the observed V2 intensities are equal to those at V1. At least one-half of the difference between the calculated and observed intensities between V1 and V2 can be explained if the heliosphere is squashed by ~10% in distance (non-spherical) so that the HTS location is closer to the Sun in the direction of V2 compared to V1.

## Introduction

The intensity recovery of galactic cosmic rays at the Earth in the current solar 11-year cycle between 2004-2009 is well documented using spacecraft data (e.g., McDonald, Webber and Reames, 2010; Mewaldt, et al., 2009, 2010). This cosmic ray recovery started in early 2004 at the Earth after the large "Halloween" events in October-November, 2003, and has been observed by neutron monitors and various spacecraft near the Earth including ACE, IMP and others. This recovery was observed by V2 and V1 to begin in the outer heliosphere in late 2004 after the Halloween event had propagated out to their respective locations at 76 and 93 AU (McDonald, et al., 2006). At the end of 2004 V1 crossed the Heliospheric Termination Shock (HTS) at 94 AU and has continued to move outward so that by 2010.5 it was at ~114 AU, perhaps ~30 AU or more beyond the current HTS location, estimated to be between 80-85 AU (Webber, 2011). Thus V1 has spent essentially the entire recovery cycle beyond the HTS in the heliosheath region where the solar wind parameters are measurably different from those in the inner heliosphere. V2 remained in the "inner" part of the heliosphere, ~15 AU closer to the sun until 2007.66 when at a distance of ~84 AU it also crossed the HTS.

At about 2010.0 the cosmic ray Helium nuclei intensity at the Earth reached its maximum. At V1 the intensity continues to increase as of 2010.5 whereas at V2 it reached a maximum in early 2009. At the Earth the intensities reached levels ~25% higher than those observed during the previous 11-year intensity maximum in 1997-98 (McDonald, Webber and Reames, 2010; Mewaldt, et al., 2009, 2010). At V1 the cosmic ray Helium intensities are at the highest levels yet observed in the heliosphere and at energies ~200 MeV/nuc at 2010.5 are within ~10-20% of the estimated LIS intensity for Helium nuclei at this energy (see Webber and Higbie, 2009).

It is the purpose of this paper to compare the Helium intensities between 150-250 MeV/nuc observed at the Earth and those observed at V1 and V2 during this extended time period within the framework of a simple modulation model, with the objective of understanding better the global characteristics of the solar 11-year modulation cycle, including particularly the modulation effects beyond the HTS in the heliosheath.

We anticipate that this is the first of several articles dealing with the recovery of cosmic ray intensities at V1, V2 and the Earth during this extended time period. Other articles will include 150-250 MeV protons and 20-125 MeV/nuc Carbon nuclei also measured at these locations during this time period. Each type of particle gives its own specific information about the heliosphere modulation process and the required "source" spectrum of the particles involved.

**Observations at the Earth and at V1 and V2**

In Figure 1 we show the time history of ~130-250 MeV/nuc He nuclei at the Earth from 2004 to the present time. This data is smoothed by taking 5 times 26 day moving averages. The data at the Earth is a composite of IMP and ACE data as constructed by McDonald, Webber and Reames, 2010. Also shown in this Figure are the corresponding intensities for ~155-245 MeV/nuc He nuclei at V1 and V2 corrected for a background of low energy ACR He (~10% or less at these energies). At the beginning of the recovery time period the intensity at V1 was ~6 times that at the Earth. This is a measure of the overall interplanetary gradient between 1 and ~94 AU, the location of V1 at that time. By 2010.5 this intensity ratio is reduced to ~2.2 implying that the intensity changes between 2004 and 2010.5 at the Earth are much greater than those at V1. This changing intensity ratio is shown in Figure 2. In Figure 3 we show the data at the Earth superimposed on the data at V1 (with different intensity scales), with the data at Earth delayed to account for the solar wind propagation time from the Earth to V1. This delay time is varied from 0.5 to 1.5 years in 26 day increments and the correlation coefficient reaches a maximum value 0.922 for time delays between 0.86 and 0.93 years. This correspondence of time histories is remarkable considering the ~100 AU difference in the radial location of the spacecraft.

This correlation throughout the heliosphere is also evident in Figure 4A which shows the intensities at V1 and V2 vs. those at the Earth, with a delay ~0.89 yrs. The "loop" in the regression curves between V1 and V2 and the Earth data in Figure 4A is due to the largest transient cosmic ray decrease in solar cycle #23 (the September, 2005, event at the Earth) propagating outward through the heliosphere, reaching V2 at ~2006.15 and V1 at about 2006.5. If this time period is excluded from the correlation calculation, the maximum value for the correlation coefficient between V1 and the Earth intensities increases to 0.961 for a delay of 0.89 years.

We seek to fit the data in Figure 4A and to interpret it using a simple global modulation model. This model should predict the absolute intensities at all three locations and also the changing ratios of intensities at V1 beyond the HTS, at V2 mainly just inside the HTS, and those intensities at the Earth vs. time as given by Figure 2 and also Figure 4A. In addition the slopes of the regression lines between V1 and V2 and the Earth, that is the ratio of the rates of change of intensity at each location needs to be fit. A simple inspection of Figures 1 and 4A shows that the intensity changes at the Earth are much larger than those at V1or V2 even though the particle energies are nearly the same.

From Figure 1 we observe that the He intensities at V2 were nearly the same as those at V1 during the minimum modulation period from 1998 to the middle of 2000. Then with increased modulation a sustained radial gradient was established between the two spacecraft which continued until after the large transient decrease in 2006 noted above, passed V2 and then V1. From early 2007 to early in 2009 the intensities at both spacecraft were almost identical again. Early in 2009 the intensity at V2 stopped increasing and by 2010.5 the difference in V1 and V2 intensities was ~20% implying again a sustained radial gradient between the two spacecraft.

**<u>The Cosmic Ray Transport Equation in the Heliosphere</u>**

Here we use a simple spherically symmetric quasisteady state no-drift transport model for cosmic rays in the heliosphere. While this simplified model obviously cannot fit all types of observations it does provide a useful insight into the inner heliospheric/outer heliospheric modulation and helps to determine which aspects of this modulation need more sophisticated models for their explanation such as a recent multi-dimensional model by Florinski and Pogorelev, 2009. The numerical model was originally provided to us by Moraal (2003) and is similar to the model described originally in Reinecke, Moraal and McDonald, 1993, and in Caballero-Lopez and Moraal (2004), and also to the spherically symmetric transport model described by Jokipii, Kota and Merenyi, 1993 (Figure 3 of that paper). The basic cosmic ray transport equation used is (Gleeson and Urch, 1971);

$$\frac{\partial f}{\partial t} + \nabla \cdot (CVf - K \cdot \nabla f) + \frac{1}{3\mathrm{p}^2}\frac{\partial}{\partial p}(p^2 V \cdot \nabla f) = Q$$

Here f is the cosmic ray distribution function, p is momentum, V is the solar wind velocity, K(r,p,t) is the diffusion tensor, Q is a source term and C is the so called Compton-Getting coefficient.

For spherical symmetry (and considering latitude effects to be unimportant for this calculation) the diffusion tensor becomes a single radial coefficient $K_{rr}$. We assume that this coefficient is separable in the form $K_{rr}(r,P) = \beta\, K_1(P)\, K_2(r)$, where the rigidity part, $K_1(P) \equiv$ K1 and radial part, $K_2(r) \equiv K2$. The rigidity dependence of K(P) is assumed to be ~P above a low rigidity limit $P_B$. The units of the coefficient $K_{rr}$ are in terms of the solar wind speed $V=4 \cdot 10^2$ $km \cdot s^{-1}$, and distance in AU = $1.5 \cdot 10^8$ km, so $K_{rr} = 6 \cdot 10^{20}$ $cm^2 \cdot s^{-1}$ when K1 = 1.0.

We consider two possible scenarios. The first is a simple heliosphere with the diffusion coefficient varying out to some outer boundary r, here taken to be 120 (130) AU, and the solar wind speed V, = const = 400 $km \cdot s^{-1}$. This is a one zone heliosphere first described by Parker, 1965. The second scenario is a two zone heliosphere (e.g., Jokipii, Kota and Merenyi, 1993). In this case the inner zone extends out to 90 AU, the average distance to the HTS. In this inner region V=400 $km \cdot s^{-1}$ and the diffusion parameters K1 and K2 are determined in our approach by a fit to the cosmic ray data being compared (the Earth and V2) rather than using e.g., consensus values (Palmer, 1982) appropriate to the "local" heliosphere.

The outer zone extends from 90 AU to ~120 (130) AU, the approximate distance to the heliopause (HP) or an equivalent "outer boundary" and essentially encompasses the heliosheath. In this region V is taken to be 130 $km \cdot s^{-1}$ (from V2 measurements, Richardson, et al., 2008) and the diffusion parameters are K1H and K2H, which are different from those in the inner heliosphere, and again determined by the cosmic ray intensity changes at V1. The distance to the HP and the source spectrum are important in this calculation.

For the LIS Helium spectrum we use the recent spectrum of Webber and Higbie, 2009. This spectrum can be approximated to an accuracy ~few % for energies above ~100 MeV/nuc by

$$\text{Helium FLIS} = (0.99/T^{2.77})/\ (1+4.14/T^{1.09}+0.65/T^{2.79}+0.0074/T^{4.20})$$

where T is in GeV/nuc. At the average energy of 200 MeV/nuc, this equation gives an input intensity of 0.98 ±0.05 $p/m^2 \cdot sr \cdot s \cdot MeV/nuc$ at the boundary at 120 (130) AU. The V1 intensity (at 114 AU) measured at 2010.5 is 0.83 in the same units, about 15% lower than the IS intensity. The intensity at V2 at the same equivalent time (+0.21 year) is 0.70 and the intensity at the Earth ~0.89 year earlier is 0.380 in the same units.

Consider a simple heliosphere with a single boundary at 120 or 130 AU. The 1st step in this approach is to fit the measured intensity at the Earth which is 0.380 at 2009.6. For K2=0 (no radial dependence of K) this requires values of K1 = 150 (165), respectively, for the two boundary locations. These values for K1 correspond, for each boundary location, to a modulation potential = 265 MV in the equivalent force field approximation where the modulation potential is defined as

$$\phi = \int_1^{R_B} \frac{V\,dr}{3K1}$$

(see Caballero-Lopez and Moraal, 2004).

This modulation potential is much lower than the average value of ~400-500 MV observed at previous sunspot minima in the modern era from 1950 (see e.g., Webber and Higbie 2010), in keeping with the unusually high intensities observed at this time in 2009 (McDonald, Webber and Reames, 2010; Mewaldt, et al., 2010). In fact the low modulation potential that we now find (based on He nuclei) is very similar to the modulation potential obtained by Mewaldt, et al., 2010, using ACE measurements of C and Fe nuclei at the Earth at the same time in 2009.

For the values of K1 which fit the data at the Earth between 2005 and 2010, however, the calculated intensities at V1 and at V2 do not provide a good fit to the data lines Figure 4A in a simple 1 zone model. If the value of K is assumed to increase with r rather than be a constant, for example, K~r, the fit to the data lines in Figure 4A is still unsatisfactory. So it is clear that a simple one zone heliosphere cannot accurately determine the intensities simultaneously observed at V1, V2 and the Earth.

For a two zone model based on an inner heliosphere inside the HTS and an outer heliosphere (the heliosheath) between the HTS and the HP with the inner heliosphere boundary at the HTS (taken here to be at 90 AU) and the HP at 120 (130) AU, we find that, for values of the HP = 120 (130) AU, values of K1=175 (max) and 42 (min) and K2=0 with V=1.0 in the inner heliosphere and values of K1H between 18 (30) (max) and 10 (24) (min), and K2H=0 with V=0.33 in the heliosheath; the two zone model accurately fits the data at the Earth and at V1 to within ± 3% over the entire time interval from 2004 to 2010 shown in Figure 4A.

As we systematically vary the values of K1 and K1H in order to fit the observed regression curves between the intensities at the Earth and V1 and the Earth and V2 during this time interval, we obtain the black and red lines shown at the time varying distances of V1 and V2 in Figure 4B. This fitting process thus provides a template as shown by these lines. This template can be moved up or down or to the left or to the right to fit the observed regression curves between the Earth data and V1 and V2 data in Figure 4A. Changes in the LIS intensity and in K1H move this template up or down and changes in K1 move it to the left or to the right. The ratio of the changes in K1 and K1H in the inner and outer heliosphere determine the slope of the black lines (these measurements are not sensitive to changes in K2). The vertical distances between the V1 and V2 lines provide a continuous measure of the effective radial intensity gradient between these two spacecraft.

These calculated intensities at the time varying distances to V1 and V2 are shown in Figure 4C using a boundary at 120 AU along with the V1 and V2 data. The predictions of the model give an overall average very good fit to the V1 intensity recovery during this 6 year time period. The predicted V1 "line" lies an average of 2% above the data and none of the smoothed data points lie more than ±10% from the predicted line. The passage of transient structures, the largest of which occurs at 2006.15 at V2, modify the overall simple sphericity of the heliosphere.

For V2 the fit is less good and the calculated He intensities are an average ~25% less than those observed, which are at times equal to those observed at V1. If the N-S asymmetry of the heliosphere, which is known to be ~10% (see Washimi, et al., 2007; Opher, et al., 2009) is taken into account, then the "effective" distance of V2 should be increased by about 10 AU and the calculated intensities at V2 should be increased by ~10-15%. This compensates for a boundary shape that is squashed in the sunward direction at V2. This improves the fit between calculations and data considerably as seen by the dashed line in Figure 4C, but the time periods of essentially zero radial gradients between V1 and V2 in 1998-99 and 2007-2008 still require additional N-S asymmetries that are time variable and mainly in the heliosheath for their explanation.

For a boundary at 130 AU the related fits to the data are shown in Figure 4D. The fit lines are very similar to those for 120 AU but require larger K1H values and a smaller change in K1H between maximum and minimum intensities than the 120 AU example. For values of the boundary >130 AU the fits deteriorate rapidly for the same assumed LIS intensity

With regard to the diffusion coefficients used in the calculations we show in Figure 5 the lower and upper limits of the values of K1 and K1H corresponding to the calculated minimum intensities in 2005 and the maximum intensities in 2010. The range of values for K1 (at 1 GV) from minimum to maximum intensities is from 42 to 175 and for K1H from 10 (24) -18 (30) for the different HP distances of 120 (130) AU. In this case the fractional change in the diffusion coefficient required to produce the minimum and maximum observed intensities in the inner zone is ~4.2 times and the change in diffusion coefficient in the outer zone is a factor ~1.80 (1.25). These fits take into account the fact that V1 has moved outward from 94 to 115 AU during the time of the measurement and V2 from 76 to 93 AU as shown by the heavy solid black and red lines in Figure 4B, 4C and 4D.

Thus, in summary, we have the situation where (1): The magnitude of the diffusion coefficient in the outer zone (heliosheath) is ~5-10 times smaller than that in the inner zone. But (2): During the intensity recovery from 2005-2010 the diffusion coefficient in the inner zone increases by a factor ~4.2 whereas in the outer zone this increase is only a factor ~1.80 (1.25). (3): For HP distances of 130 AU or greater, the IS He intensity must increase in order to fit the V1 data. (4): The observed V2 data is between 20 and 25% higher than the predictions during this time period. But assuming a squashed heliosphere within an asymmetry ~10% (8-10 AU at the HTS) this difference decreases to ~10% or less.

## Summary and Conclusions

The recovery of the intensity of ~150-250 MeV/nuc cosmic ray He nuclei has been followed between 2004-2010 at the Earth and also at V1 and V2 in the outer heliosphere and in the case of V1, beyond the HTS. The correlation of the intensity changes at the Earth and V1 in the outer heliosphere (correlation coefficient =0.922), ~100 AU apart, is remarkable after accounting for a time delay ~0.9 year due to the solar wind propagation. The relative intensities at V1, V2 and at the Earth as well as the slope of the regression lines between the measurements place limits on the amount of solar modulation in the inner and outer heliosphere. It is found that the data at the Earth and at V1 can be reproduced by a simple two zone heliosphere where the intensity changes are due to changes in the cosmic ray diffusion coefficient K in each zone. In the inner zone, out to the HTS assumed to be at 90 AU, the value of K is quite large (see Figure 5) and varies by a factor ~4.2 from the minimum to maximum modulation in this part of the solar 11-year cycle. In the outer zone from ~90-120 (130) AU, essentially in the heliosheath, the value of the diffusion coefficient is much smaller, by a factor ~5-10 and varies by a factor ~1.80 (1.25) from minimum to maximum modulation.

Thus, in effect, because of the small value of the diffusion coefficient, the heliosheath is a very turbulent, diffusive region, acting much like a diffusive barrier to these lower energy cosmic rays in spite of the slower solar wind speed and other effects which tend to greatly reduce the effects of adiabatic energy loss beyond the HTS.

Although the V1 Helium data is fit to a level ~±5% over the entire time period from 2004-2010 for boundaries between 120-130 AU, the V2 Helium data is not well fit with a simple spherically symmetric heliosphere, with the predicted intensities typically ~25% less than the data. If the heliosphere in the V2 direction is assumed to be flattened in the sunward direction with an asymmetry ratio as determined by Washimi, et al., 2007, see also Opher, et al., 2009, then the model fit to the V2 data is generally better (the differences between predictions and observations are now ~10% or less), but the fact that there are extended periods of essentially zero radial gradient between V1 and V2 require times of additional time variable asymmetries between the N and S hemispheres, mainly taking place in the heliosheath.

The details of the fit to the data beyond the HTS depend on the values of the local interstellar spectrum (LIS) used as an input to the modulation calculation and also the location of the heliopause or boundary to the modulation region. For the estimated LIS He intensity used in this paper the V1 data can be well fit for HP distances in the range of 120-130 AU. This heliosheath region and the interstellar helium spectrum itself will be mapped in more detail as V1 continues to move outward in the heliosphere and the intensity continues to increase towards the LIS value.

**Acknowledgements:** The authors wish to thank the Voyager team (E.C. Stone, P.I.) and the ACE team (R.A. Mewaldt, P.I.) for making their data available on their web-sites, http://voyager.gsfc.nasa.gov and http://www.srl.caltech.edu/ACE/.

## References

Caballero-Lopez, R.A. and H. Moraal (2004), Limitations of the force field equation to describe cosmic ray modulation, J. Geophys. Res., 109, A01101, doi:10.1029/2003JA010098

Florinski, V. and N.V. Pogorelev, (2009), Four dimensional transport of galactic cosmic rays in the outer heliosphere and heliosheath, Ap.J., 701, 642-651

Gleeson, L.J. and I.A. Urch, (1971), Energy losses and the modulation of galactic cosmic rays, Astrophys. Space Sci., 11, 288-308

Jokipii, J.R., J. Kota and Merenyi, (1993), The gradient of galactic cosmic rays at the solar wind termination shock, Ap. J., 405, 753-786

McDonald, F.B., W.R. Webber, E.C. Stone, A.C. Cummings, B.C. Heikkila and N. Lal, (2006), Voyager observations of galactic and anomalous cosmic rays in the Heliosheath, AIP Conf. Proc., 858, 79-85, doi:10.106311.2359309

McDonald, F.B., W.R. Webber and D.V. Reames, (2010), Unusual time histories of galactic and anomalous cosmic rays over the deep solar minimum of cycle 23/24, Geophys. Res. Lett., 37, L18101, doi:10.1029/2010GL044218

Mewaldt, R.A., R. Leske and K. Lave, (2009), Cosmic ray Fe intensity reaches record levels in 2008-2009, ACE News #122, http://www.srl.caltech.edu/ACE/ACENews/ACENews122.html

Mewaldt, R.A., et al., (2010) Record-setting cosmic ray intensities in 2009 and 2010, Ap.J. Lett., 723, L1-L6

Opher, M., J.D. Richardson, G. Toth and T.I. Gombosi, (2009), Confronting Observations and Modeling: The role of the Interstellar Magnetic Field in Voyager 1 and 2 Asymmetries, Space Sci. Rev., 143, 48-55

Palmer, I.D. (1982), Transport coefficient of low-energy cosmic rays in the interplanetary space, Rev. Geophys. Space Phys., 20, 335

Parker, E.N., (1963), Interplanetary dynamical process, New York, Interscience

Reinecke, J.P.L., H. Moraal and F. B. McDonald, (1993), The cosmic radiation in the heliosphere at successive solar minima: Steady state no-drift solutions of the transport equations, J. Geophys. Res., 98, (A6), 9417-9431

Richardson, J.D., J.C. Kasper, C. Wang, J.W. Belcher and A.J. Lazarus, (2008), Cool heliosheath plasma and deceleration of the upstream solar wind at the termination shock, Nature, 454, 63-66, doi:10.1038/nature07024

Washimi, H., G.R. Zank, Q. Ha, T. Tanaka and K. Munakata, (2007), A forecast of the heliospheric termination shock position by three dimensional MHD stimulation, Ap.J., 670, L139-L142

Washimi, H., G.R. Zank, Q. Ha, T. Tanaka and K. Munakata, (2010), Realistic and time-Varying Outer Heliospheric Modeling by Three-Dimensional MHD Simulation, Ap.J., (in press)

Webber, W. R. and P. R. Higbie, (2009), Galactic propagation of cosmic ray nuclei in a model with an increasing diffusion coefficient at low rigidities: A comparison of the new interstellar spectra with Voyager data in the outer heliosphere, J. Geophys. Res., 114, A02103, doi:10.1029/2008JA013689.

Webber, W.R., and P.R. Higbie, (2010), What Voyager cosmic ray data in the outer heliosphere tells us about 10Be production in the Earths polar atmosphere in the recent past, J. Geophys. Res., 115, A05102, doi:10.1029/2009JA014512

## Figure Captions

**Figure 1:** 5 x 26-day running average of V1, V2 and IMP/ACE 150-250 MeV/nuc He nuclei data from 1998 to 2010.5. The Earth data is delayed by 0.89 year to account for inner-outer heliosphere delay in modulation due to solar wind propagation time.

**Figure 2:** 5 x 26 day running average of V1 to Earth ratio of 150-250 MeV/nuc He nuclei from 2004 to 2010.5 (Earth data delayed by 0.89 year).

**Figure 3:** The V1 data from 2004.8 in Figure 1 superimposed on the data at the Earth delayed by 0.89 year (with different intensity scales on the left and right axis). This figure shows the high level of correlation between intensity changes at the Earth and in the outer heliosphere during this time period.

**Figure 4A:** Regression plot of the observed intensities from 2004.8 at V1 and V2 vs. the intensities at the Earth delayed by 0.89 year. Both axis in Figures 4A, 4B, 4C and 4D are $P/m^2 \cdot s \cdot sr \cdot MeV/nuc$

**Figure 4B:** Solid black (V1) and red (V2) lines show the predictions of the He nuclei intensities along the V1 and V2 trajectory in a two-zone heliospheric modulation model with the boundary at 120 AU.

**Figure 4C:** The V1 (black) and V2 (red) data points superimposed on the model predictions of Figure 4B, ($R_B$ = 120 AU), K1H = 18 (max) to 10 (min). The effect of a general heliospheric radial N-S asymmetry ~10% near the HTS on the predictions for V2 is shown as a dashed line.

**Figure 4D:** Same as Figure 4C but with $R_B$ = 130 AU and K1H changing from 30 (max) to 24 (min).

**Figure 5:** Values of K1 and K1H used in the two-zone modulation model. Black lines labeled 2010 and 2005 show the range of values of K in the inner heliosphere and in the heliosheath that are necessary to reproduce the He intensity changes observed between 2005 and 2010.5 at the Earth and at V1 and V2. The solid points at 1 GV indicate the maximum and minimum values of K1 and K1H at that rigidity.

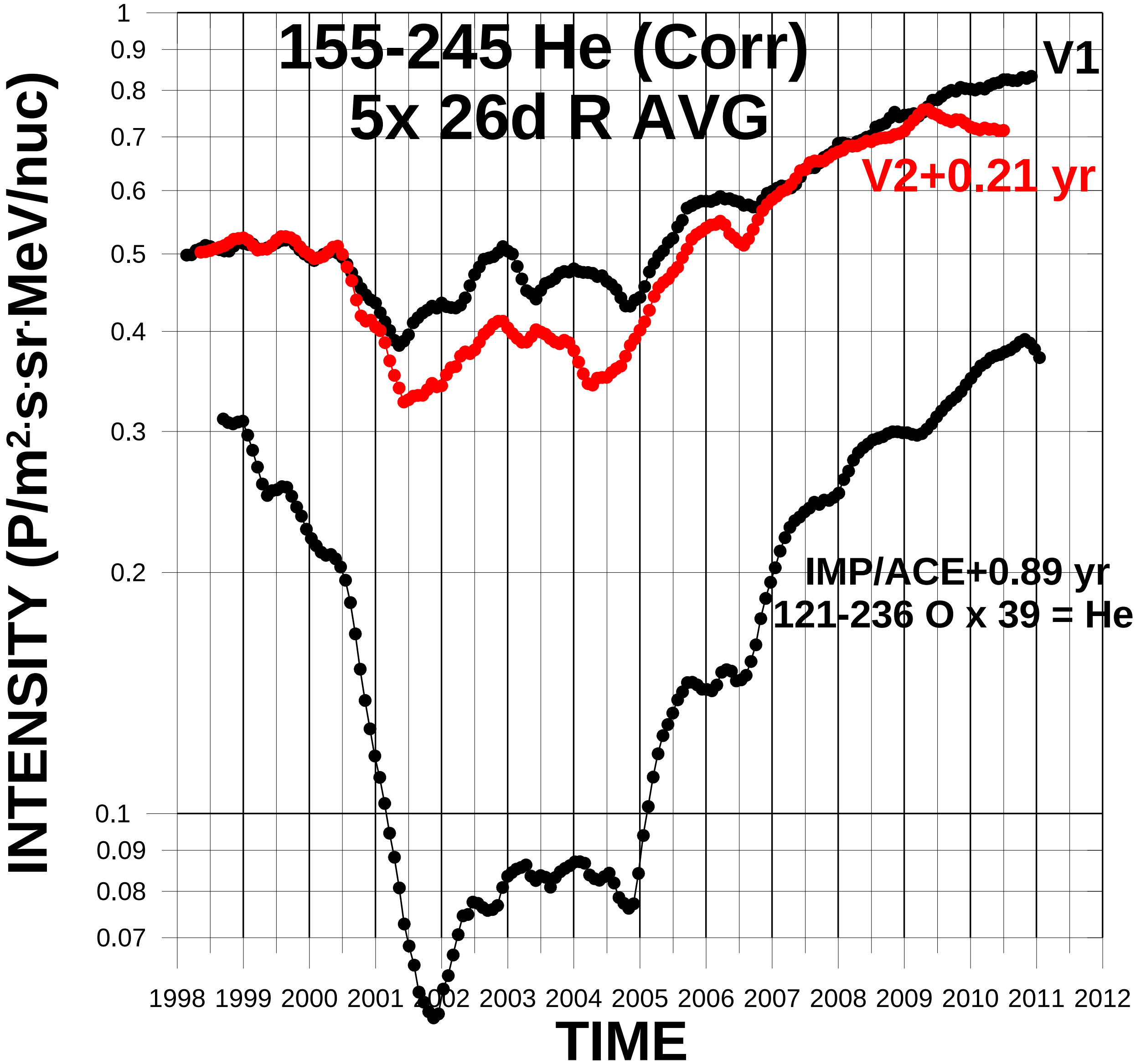

346 **FIGURE 1**

347

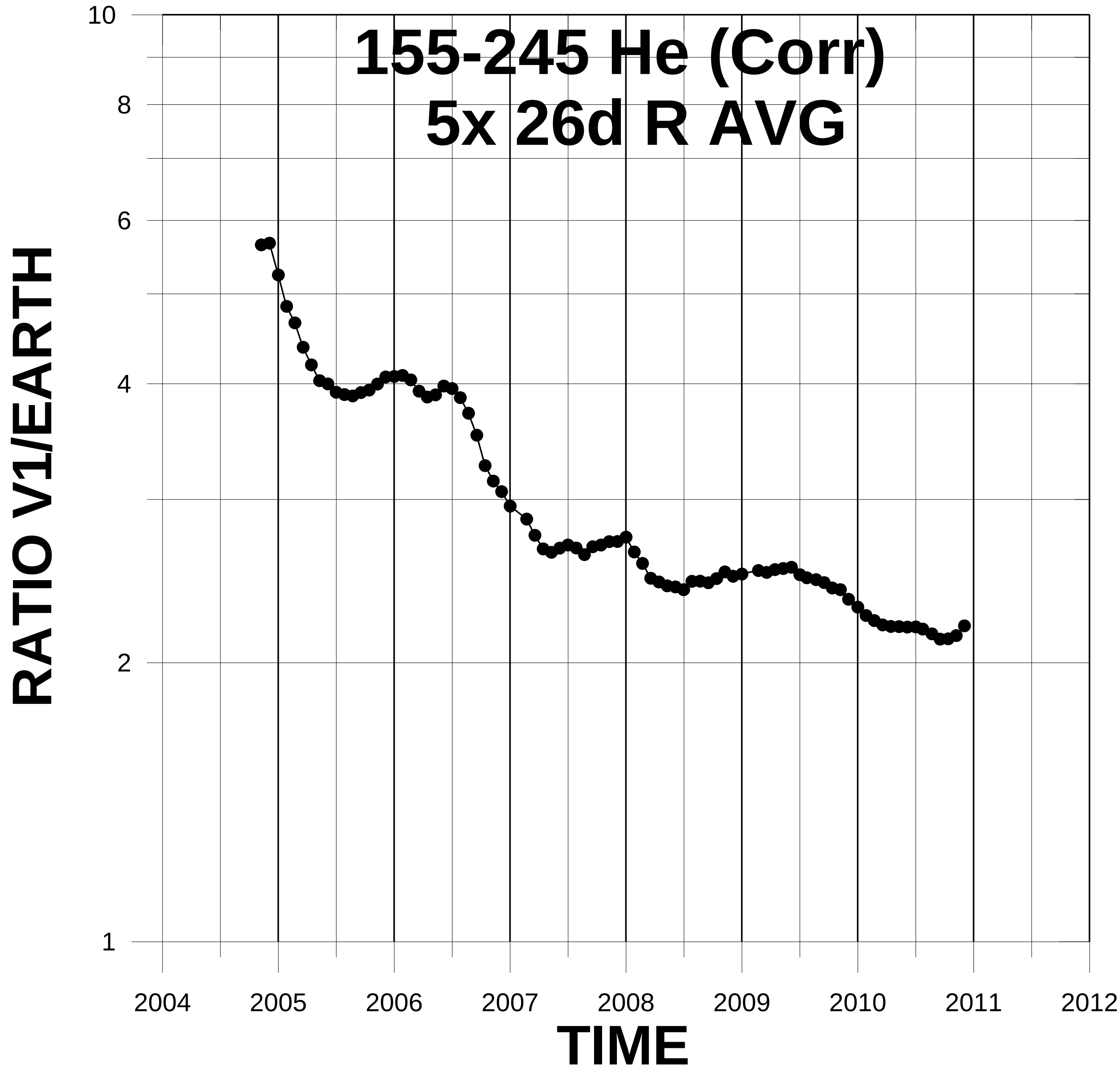
155-245 He (Corr)
5x 26d R AVG
RATIO V1/EARTH
10
8
6
4
2
1
2004
2005
2006
2007
2008
2009
2010
2011
2012
TIME

**FIGURE 2**

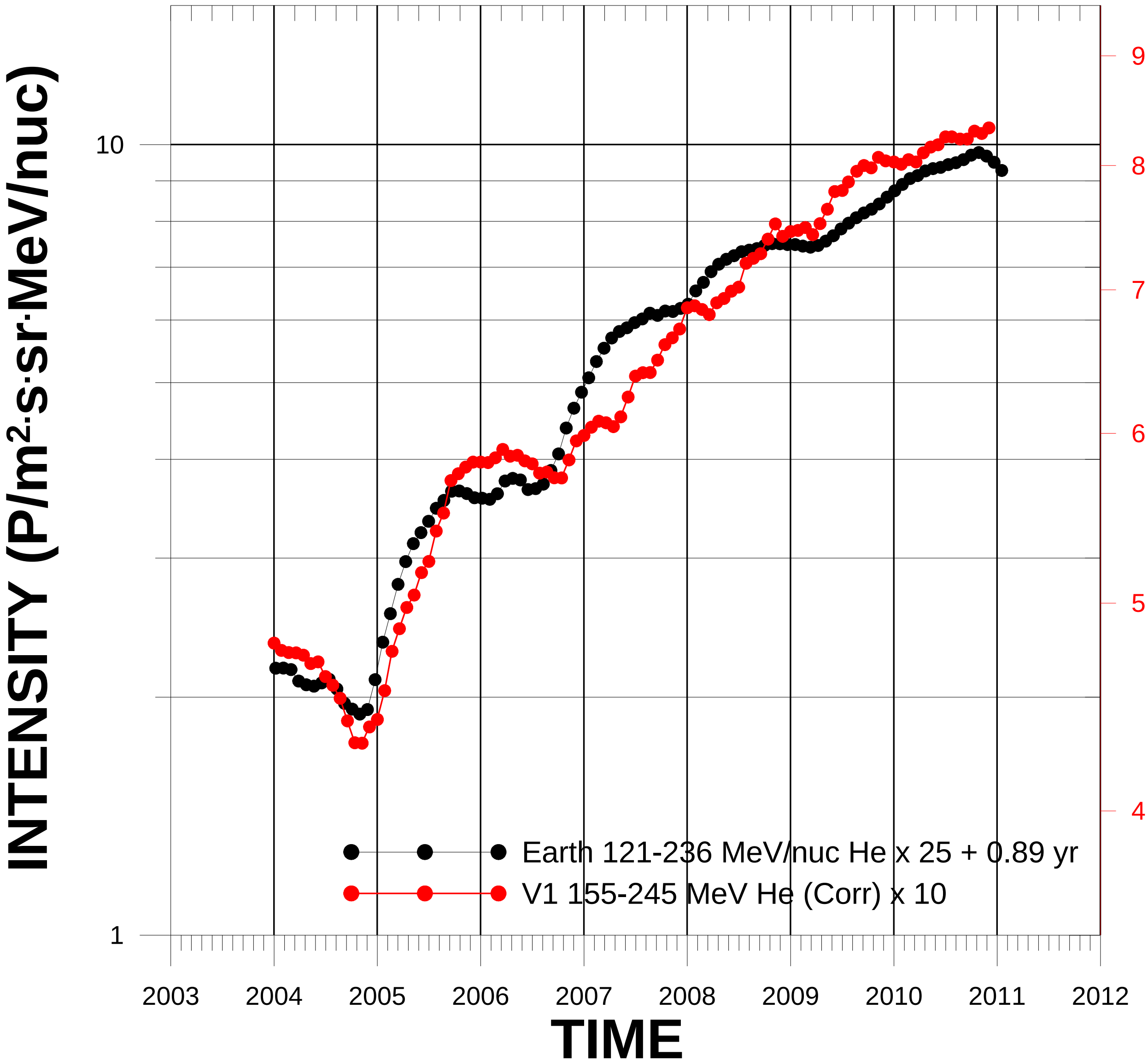

INTENSITY (P/m²·s·sr·MeV/nuc)
10
1
9
8
7
6
5
4
Earth 121-236 MeV/nuc He x 25 + 0.89 yr
V1 155-245 MeV He (Corr) x 10
2003
2004
2005
2006
2007
2008
2009
2010
2011
2012
TIME

**FIGURE 3**

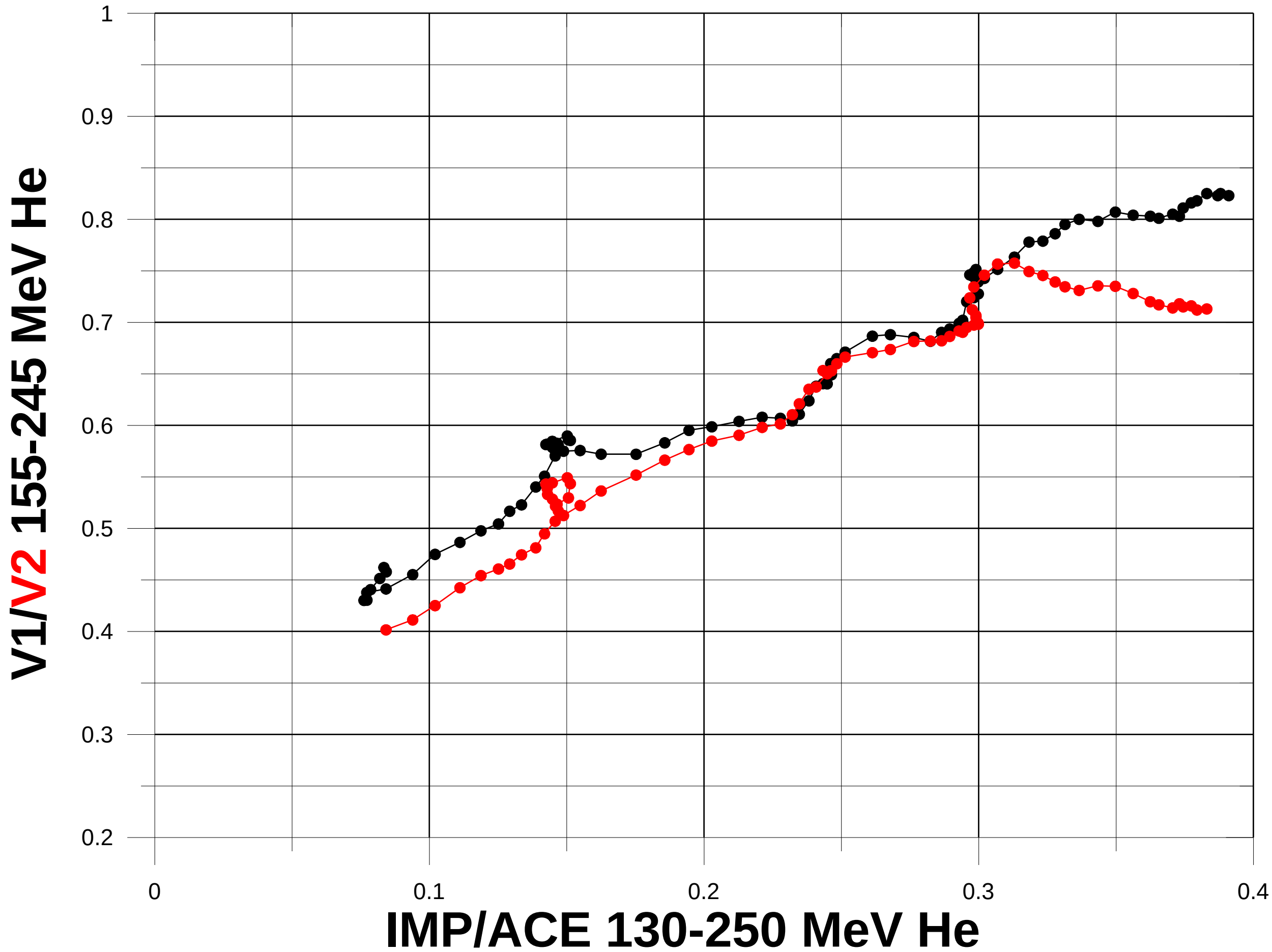
1
0.9
0.8
0.7
0.6
0.5
0.4
0.3
0.2
V1/V2 155-245 MeV He
0
0.1
0.2
0.3
0.4
IMP/ACE 130-250 MeV He

**FIGURE 4A**

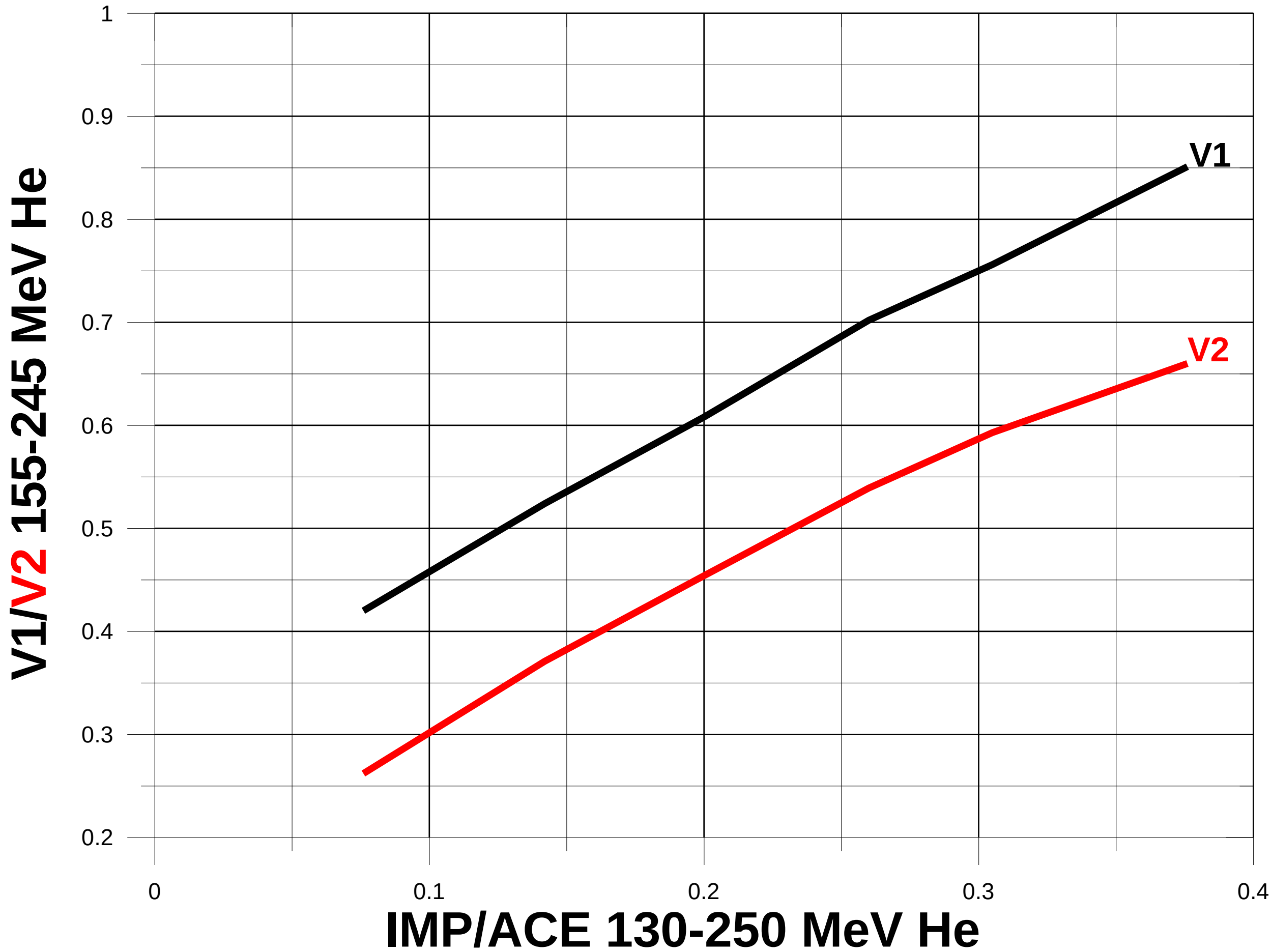
V1/V2 155-245 MeV He
1
0.9
0.8
0.7
0.6
0.5
0.4
0.3
0.2
V1
V2
0
0.1
0.2
0.3
0.4
IMP/ACE 130-250 MeV He

**FIGURE 4B**

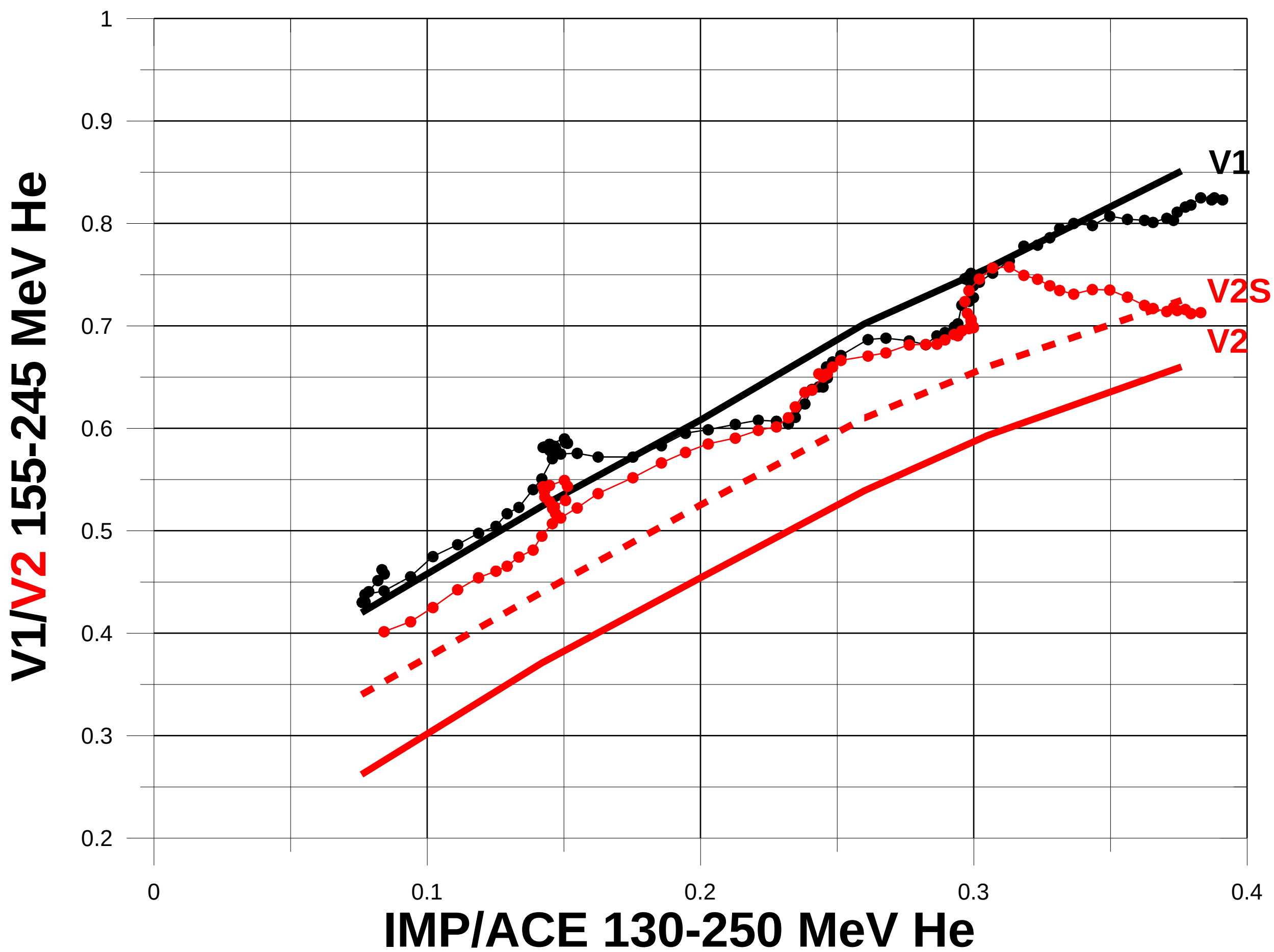

356 **FIGURE 4C**

357

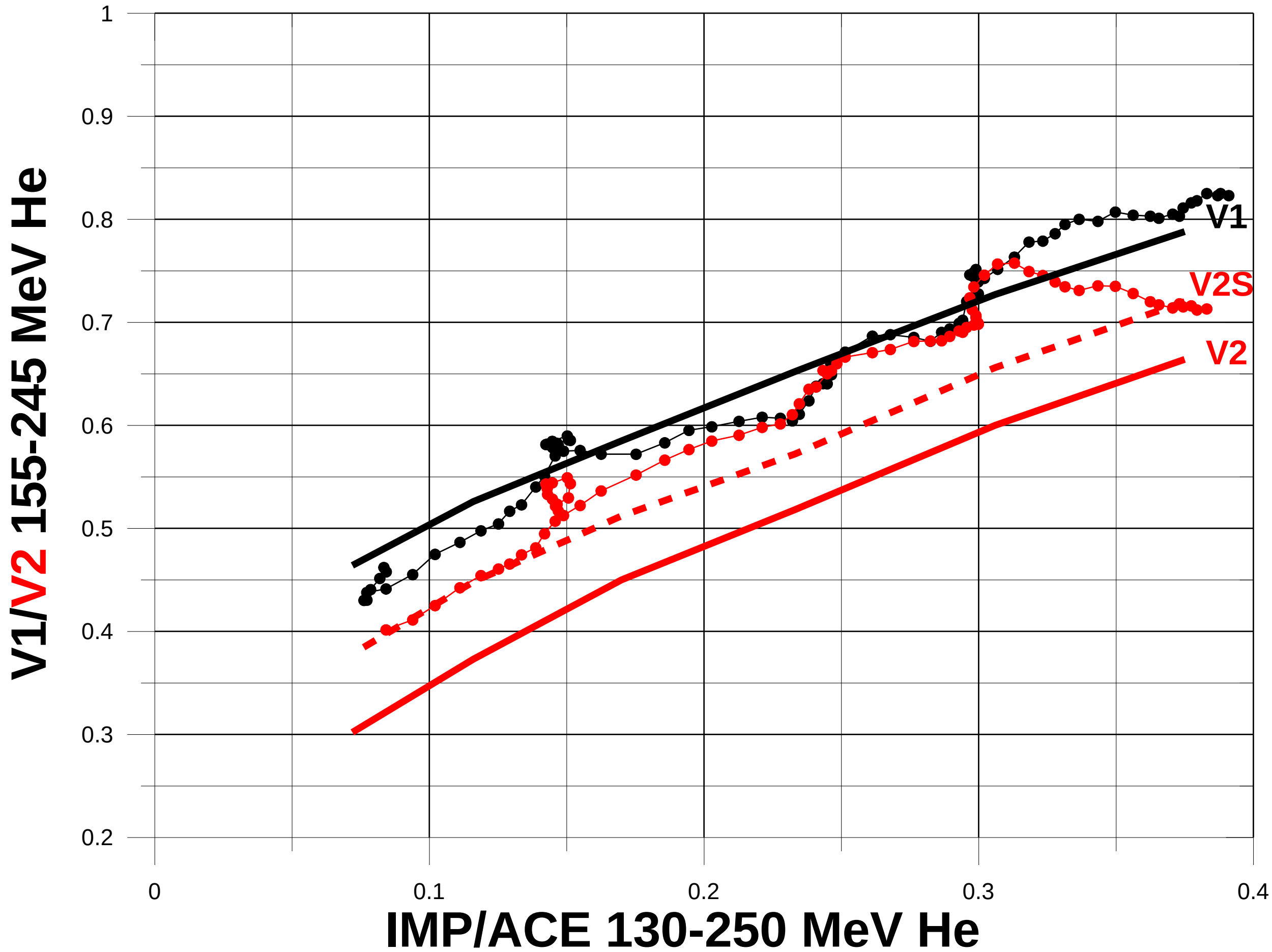
V1/V2 155-245 MeV He
IMP/ACE 130-250 MeV He
V1
V2S
V2
1
0.9
0.8
0.7
0.6
0.5
0.4
0.3
0.2
0
0.1
0.2
0.3
0.4

FIGURE 4D

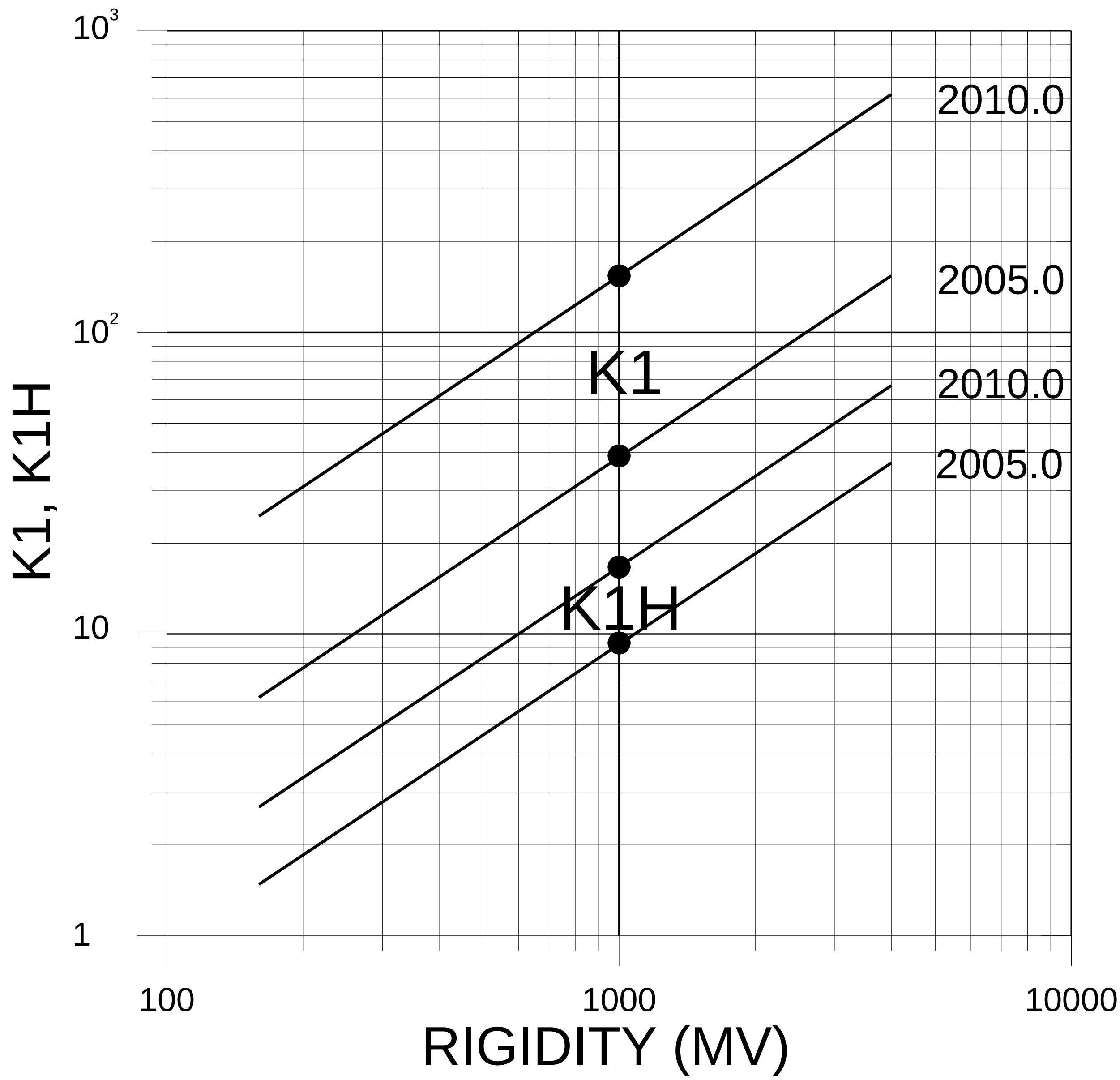

10^3
10^2
10
1
2010.0
2005.0
2010.0
2005.0
K1
K1H
K1, K1H
100
1000
10000
RIGIDITY (MV)

FIGURE 5